\documentstyle[12pt]{article}

\newcommand{\algebra}{{\cal A}}
\newcommand{\algspan}{{\rm span}}
\newcommand{\area}{{{\cal O}_{\lambda}}}

\newcommand{\hh}{{\cal H}}

\newcommand{\obs}{{\cal O}}
\newcommand{\ooo}{{\bf O}}
\newcommand{\pq}{{|p\!><\!q|}}

\newcommand{\rota}{\rho}

\newcommand{\tends}{\rightharpoonup}

\date{}
\title{Topology measurement within the histories approach}

\author{
R.Breslav\footnotemark[1],
G.N.Parfionov\footnotemark[2],
R.R.Zapatrin\footnotemark[2]{\makebox[0.4em]{}}\footnotemark[3]
}

\begin{document}

\raggedbottom
\topmargin-49pt
\textheight620pt

\maketitle
\addtocounter{footnote}{+1}
\footnotetext{The Centre of Mathematics, Nevsky pr., 39,
191011, St-Petersburg, Russia}
\footnotetext{Department of Mathematics, SPb UEF, Griboyedova 30/32,
191023, St-Petersburg, Russia (address for correspondence)}
\addtocounter{footnote}{+1}
\footnotetext{Division of Mathematics,
Istituto per la Ricerca di Base,
I-86075, Monteroduni (IS), Molise, Italy}

\begin{abstract}
An idealised experiment estimating the spacetime topology is
considered in both classical and quantum frameworks. The latter is
described in terms of histories approach to quantum theory. A
procedure creating combinatorial models of topology is suggested.
The correspondence between these models and discretised spacetime
models is established.
\end{abstract}

\newpage

\section{Introduction}

Within the conventional account of the relativity theory the
structure of spacetime as differentiable manifold is supposed to be
given and it is the metric structure that is subject to measurement
and changes. So, the topology of spacetime is not {\em an
observable.}

Nowadays there is no fully fledged theory in which the spacetime
topology would be a variable, nor even in a sense perceivable
entity. However, even if such theory does not exist, we may try to
consider idealized experiments which would let us know the
spacetime topology. That means, we should assume the spacetime
to be a manifold, and we only wish to {\em determine} its
topological structure. In accordance with it, any observer should
believe that the topology of the area of his observation (that is,
appropriate coordinate neighborhood) is that of a ball. So, in
order to recover the entire spacetime topology we have to find out
how the balls do overlap. However, any realistic experiment (having
at most finite number of outcomes) can not let us know it. We are
only able to know if the regions have common points (section
\ref{semp}).

Such experimental scheme inevitably needs {\em several} observers,
but the problem of event identification arises: two observers
registering an event should be made sure they really see the same.
We emphasize that this is a matter of {\em convention:\/} two
observers should have a way to identify remote events. This leads
to the concept of organized observation (section \ref{sentang}).

The obtained results of observations then ought to be somehow
interpreted. We may do that in classical either quantum way. In the
classical approach this leads to the Sorkin discretization scheme
(section \ref{sfinsub}).

The attempt to put a scheme of topology estimation into the
framework of quantum mechanics requires the cooperative nature of
the observations to be explicitly captured in the theory. It is the
notion of {\em homogeneous history} in the histories approach to
quantum theory \cite{ha} that can be used for this purpose. Within
the histories approach we introduce the notion of the
'team' (organized set of observers). To carry out the mathematical
description of the team we had to impose an additional mathematical
structure. It turned out that this structure can be represented by
that of associative algebra (section \ref{sentang}). It is worthy
to mention that such structures can be introduced in different ways
reflecting different ways of organization of the team of observers.

\[
\left( \begin{array}{c}
\hbox{topology} \cr
\hbox{measurement}
\end{array} \right)
\,=\,
\left( \begin{array}{c}
\hbox{homogeneous} \cr
\hbox{history}
\end{array} \right)
\,+\,
\left( \begin{array}{c}
\hbox{organization} \cr
\hbox{of observers}
\end{array} \right)
\]

There is no spacetime points at all within the histories approach,
and the goal of the introduced additional structure is to
'manufacture' them. We suggest an algebraic machinery building
topological spaces (namely, the Rota topologies on primitive
spectra of appropriate algebras) and call it {\em spatialization
procedure} (section \ref{sspat}).

In order to make sure of the viability of our quantum construction
we should take care of {\em correspondence principle}: we should be
able to carry out quasiclassical measurements. That means to
possess such an organization scheme for the team that the result of
the spatialization procedure would be the same as in classical
approach. It reduces to a purely mathematical problem of existence
of appropriate algebraic structure. We suggest a constructive
solution of this problem using so-called incidence algebras
(section \ref{sialg}).

\section{Empirical topology}\label{semp}

Let us consider, following \cite{g2}, an idealized experimental
scheme for determining the topological structure of spacetime.
Consider a team $\Lambda$ of observers. Each of them assumes
himself to be in the center of an area $\area$ $(\lambda \in
\Lambda$) homeomorphic to an open ball. We require it to satisfy
the correspondence principle:  in fact, looking around we do not
see holes or borders in the sky.  These areas $\{\area\}$ will form
an atlas for the spacetime manifold in which they are. Then the
problem of learning the structure of the entire manifold arises. It
was solved by Alexandrov \cite{alexandrov} by introducing the
notion of {\em nerve} of the covering, namely the result is encoded
in the structure of mutual intersections of the elements of the
covering.

\medskip

Within the proposed scheme, the problem is to {\em experimentally}
verify which areas $\area$ do overlap. This is done by exchanging
information between observers about the events they observe. The
results of the observations could be put into the following table
(Tab. \ref{tab1}) whose rows correspond to events and columns
correspond to the observers.

\begin{table}[h!t]
\begin{center}
\begin{tabular}{||c||c|c|@{$\:\ldots\:$}|c|@{$\:\ldots\:$}||}
\hline
Event label & $\obs_1$ & $\obs_2$ & $\obs_\lambda$ \cr
\hline
1 & + & -- & + \cr
2 & + & + & + \cr
\ldots & \ldots & \ldots & \ldots  \cr
$n$ & -- & + & +  \cr
\ldots & \ldots & \ldots & \ldots  \cr
\hline
\end{tabular}
\end{center}
\caption{The results of observations: if an observer $\lambda$
registers the event $i$ we put "$+$" into the appropriate
cell of the table, otherwise "$-$" is put.}
\label{tab1}
\end{table}

The consequences we make out of the experiments necessarily
have the statistical nature. In particular, the statement "the
areas of two observers do overlap" is merely a statistical
hypothesis. To verify it the following criterion is suggested:

\begin{equation}\label{epropos}
\mbox{\parbox[c]{100mm}{\em
If it occurs that the observers $\obs_1$ and $\obs_2$ have
registered the same event, then the areas of their observations do
overlap.
} }
\end{equation}

Note that this criterion is {\em statistical} rather than {\em
logical}. We emphasize that after the observations were carried out
we only accept or reject the appropriate hypothesis.

\medskip

When such a hypothesis is accepted, it gives us the complete
information about the {\em nerve} of the covering. One might think
that now we are able to recover the global topology by gluing the
balls together. But this is an illusion: the obstacle is that we
have nothing to glue!  Moreover the geometrical realization by
nerve may be a source of artifacts: for instance we can cover an
interval $(0,1)$ (having dimension 1) in such a way

\[
\begin{array}{rcl}
\obs_1 &=& (0,0.6) \\
\obs_2 &=& (0.4,1) \\
\obs_3 &=& (0.2,0.8)
\end{array}
\]

\noindent that the appropriate nerve is realized by a triangle
(having dimension 2). Supposed we could exhaust {\em all} the
points of spacetime, the "real ultimate" structure of spacetime
manifold would be recovered.  However what we can really carry out
is to realize a "homogeneous history" whose outcome is recorded in
the table like Tab.  \ref{tab1}.

\section{Entanglement in histories approach}\label{sentang}

In this section we introduce topology measurements into the
histories approach to quantum theory. It will be based on the
algebraization scheme of the histories approach suggested by
C.Isham \cite{qlha}. The key issues of this scheme are

\begin{itemize}
\item to consider {\em
propositions} about histories rather than histories themselves
\item to span a linear space on elementary propositions about
histories
\item to endow the propositions themselves by the additional
structure of orthoalgebra
\end{itemize}

\noindent thus organizing them in a way similar to conventional
quantum mechanics.

In this paper a similar idea is realized. We consider
\begin{itemize}
\item[*] propositions about topologies rather than the topologies
themselves
\item[*]  a linear space
spanned on the elementary propositions about topologies
\item[*]  the structure of associative algebra on this
linear space
\end{itemize}

Let us specify what do we mean by propositions about topologies.
There are at least three ways to introduce a topology on a set $M$
\cite{ishamtop}. First two of them are in a sense exhaustive: to
define (to list out) all open sets either to define the operation
of closure on all subsets of $M$. The third way is more 'economic':
to declare which sequences do converge. It will be suitable for us
to replace topology by convergencies for both technical and
operationalistic reasons (section \ref{sspat}.  In fact, any
realistic experiment can yield us at most a finite sequence of
results. The associative algebras related with propositions about
topology will be built in section \ref{sialg}.

Let us figure out how the notion of organized team of observers can
be incorporated into the histories approach. Let

\begin{equation}\label{eh1}
A^1_{t_1}U_{t_1t_2} A^2_{t_2}\ldots
U_{t_{n-1}t_n}A^n_{t_n}\psi_0
\end{equation}

\noindent be a homogeneous history. The operators $A^i$ are assumed
to act in a Hilbert space $\hh$. It was suggested by Isham
\cite{qlha} to describe the history (\ref{eh1}) by an element of
the tensor product $\otimes_{i=1,n}\hh$. Then we assume that there
is an 'organizer' of the history whose status is {\em a priori} the
same as that of every member of the team. That means that he has
the same state space $\hh$. Thus each history, that is, a vector
from $\otimes_{i=1,n}\hh$, should be associated with a vector in
the state space $\hh$ of the organizer. The suggested
correspondence should meet the following requirements:

\begin{itemize}
\item[(i)] Neither the number of observers nor their particular
choice of what to measure should influence the form of this
organization \item[(ii)] If we have an experiment which is a
refinement of different coarser experiments, their results should
not contradict \item[(iii)] This correspondence should be linear in
order to support the superposition principle \end{itemize}

\medskip

Mathematically this correspondence is introduced by defining a
family of linear mappings $\ooo_n$ ($n=1,\ldots,n$):

\begin{equation} \label{eooo}
\ooo_n :\hh \otimes \hh \otimes \ldots \otimes \hh\rightarrow \hh
\end{equation}

\noindent whose form is specified by a particular organization of
the topology measurement. The requirement (iii) is expressed in
the linearity of $\ooo_n$. To meet the requirement (i) we are
dealing with the family $\{\ooo_n\}$ rather than with a single
mapping.

Now the requirement (ii) can be formulated as a relation between
the mappings $\ooo_n$. First,

\[ \ooo_1(x) = x \]

\noindent and

\[
\ooo_{p+q}(x_1\otimes \ldots \otimes x_{p+q}) =
\ooo_2(\ooo_p(x_1\otimes \ldots \otimes x_p) \otimes
\ooo_q(x_1\otimes \ldots \otimes x_q) )
\]

\noindent In particular

\begin{equation}\label{eass}
\ooo_2(\ooo_2(x\otimes y)\otimes z) =
\ooo_2(x\otimes \ooo_2(y\otimes z))
\end{equation}

\noindent therefore all the mappings $\ooo_n$ can be inductively expressed
through $\ooo_2$.

\[
\ooo_n(x_1\otimes \ldots \otimes x_{n-1} \otimes x_n) =
\ooo_2(\ooo_{n-1}(x_1\otimes  \ldots \otimes x_{n-1}) \otimes x_n)
\]

\medskip

Being a linear mapping, $\ooo_2:\hh \otimes
\hh \to \hh$ generates a bilinear mapping $\hh \times \hh \to
\hh$ whose action on the pair $(x,y)$ we denote simply by
$x\cdot y$. Then the relation (\ref{eass}) reads:

\[
(x\cdot y)\cdot z  =
x\cdot (y\cdot z)
\]

So, the organization of a topology measurement is mathematically
expressed by defining an {\em associative} product in $\hh$. Then
the 'organizing operator' (\ref{eooo}) takes the form:

\[
\ooo_n(x\otimes y\otimes \ldots \otimes z) = x\cdot y\cdot
\ldots \cdot z
\]

Suppose a history realizing a topology measurement 'took place',
that is, the team of observers had carried out a number of yes-no
experiments (Table \ref{tab1}), or, in other words, the operators
$A^i_{t_i}$ are projectors in $\hh$:

\begin{equation}\label{eh10}
P^1_{t_1}U_{t_1t_2} P^2_{t_2}\ldots
U_{t_{n-1}t_n}P^n_{t_n}\psi_0
\end{equation}

The outcome of each of these yes-no measurements is the selection
of a subspace of $\hh$ (associated with appropriate projector). The
organizer has a collection of subspaces of $\hh$ in his disposal.
Now let us return to requirement (i):  what invariant object may he
construe out of them having only these subspaces and the product in
$\hh$? This is the algebra $\algebra$ spanned on these subspaces.
So, all the available information about the spacetime topology is
encoded in this subalgebra of $\hh$. A way to extract it is to apply
the spatialization procedure described in the next section.

\section{Spatialization procedure and Rota topology}\label{sspat}

Let us consider what sort of spaces can be extracted from algebras.
Begin with the discussion of what points ought to be. Suppose for a
moment that the obtained algebra $\algebra$ is commutative. In this
case it can be canonically represented by a functional algebra on
an appropriate topological space. This may be obtained using
Gel'fand representation. In this case the points of this space can
be thought of as characters. The characters are, in turn,
one-dimensional irreducible representations whose kernels are
maximal ideals. There are several ways to impose a topology on the
set of points \cite{bourbaki}.

In general, when the algebra $\algebra$ may be non-commutative, the
scheme of geometrization remains in principle unchanged: we only
pass from characters to classes of irreducible representations,
and, respectively, from maximal ideals to primitive ones. For a
more detailed analysis of the relevance of primitive ideals the
reader is referred to \cite{g1}. So

\[
X = {\rm Prim}\,\algebra
\]

\noindent that is, the points are the elements of the primitive
spectrum of $\algebra$ (equivalence classes of IRRs). Note that at
this point we have $X$ as a {\em set} not yet endowed by any
structure. The straightforward way to 'topologize' $X$ could be to
use the Jacobson topology. Unfortunately, in the finitary context
we are (section \ref{semp}) this topology (as well as the
other standard ones) reduces to the trivial case of discrete one.
So, let us seek for a weaker structure which could produce
us a reasonable topology on $X$.

It is the notion of convergence space \cite{ishamtop} which is the
closest to topological structure. It is formed by declaring a
relation $(x_n)\tends y$ of convergence between sequences and
points:

\[
x_1,x_2,\ldots, x_n, \ldots \tends y
\]

\noindent which always gives rise to the following relation on the
points of $X$:

\begin{equation}\label{econv}
x\tends y \:\hbox{if and only if}\:
x,x, \ldots,x, \ldots \tends  y
\end{equation}

Having any relation $\tends$ on $X$, we are always in a position to
define a topology on $X$ as the strongest topology in which
(\ref{econv}) holds. So, we shall introduce a topology on $X = {\rm
Prim}\,\algebra$ according to the following scheme:

\[
(\hbox{relation on }X) \longrightarrow (\hbox{topology on }X)
\]

Recall that the elements of $X$ are the primitive ideals of
$\algebra$, which are, in turn, subsets of $\algebra$. Having two
such ideals $X,Y$ we can form both their intersection $X\cap Y$ and
their product $X\cdot Y$ as the ideal spanned on all products
$x\cdot y$ with $x\in X$, $y\in Y$. Note that in general $X\cdot
Y\neq Y\cdot X$. However both $X\cdot Y$ and $Y\cdot X$ always lie
in (but may not coincide with) $X\cap Y$. Relations between
primitive ideals were investigated. G.-C. Rota \cite{rota}
introduced the following relation in the context of enumerative
combinatorics:

\begin{equation}\label{etends}
X \rota Y \:\hbox{if and only if}\: X\cdot Y
\stackrel{\neq}{\subset} X\cap Y
\end{equation}

We shall call the topology generated by this relation "$\rota$"
the {\sc Rota topology} on the set of primitive ideals.

\medskip

We see that in order to judge on the measured topology it suffices
to build an 'organizing' algebra. A particular form of this
algebra should be produced using the table of observations (like
Tab. \ref{tab1}). There is no {\em a priori} preferred way to build
such an algebra: different models of 'data processing' may give
different spatializations. However, there exists a 'classical
spatialization' using no quantum models --- this is the Sorkin
discretization scheme (section \ref{sfinsub}). The problem of
correspondence then arises is it possible to suggest such an
organizing algebra based on the table of results that the
appropriate topological spaces (Rota and Sorkin topologies) would
coincide. This problem will be solved in section \ref{sialg}.

\section{Finitary substitutes}\label{sfinsub}

The Sorkin spatialization procedure imposes the topology on the set
$N$ of events whose prebase is formed by the subsets of events
observed by each observer. Consider this construction in more
detail following the account suggested in \cite{g2,g1}.

Associate with any event $i$ the set $\Lambda_i \subseteq
\Lambda$ of observers which registered it:

\begin{equation}\label{e11}
\Lambda_i = \{\lambda \subseteq
\Lambda \mid \quad \hbox{the event}\; i \; \hbox{was
registered by }\, \lambda\}
\end{equation}

\noindent and consider the relation $\tends$ on the set of events:

\begin{equation}\label{e12}
i\tends j \:\:\, \hbox{if and only if} \:\:\,
\forall \lambda \: j\in N_\lambda \, \Rightarrow \,
i\in N_\lambda
\end{equation}

Note that the relation $\tends$ is evidently reflexive ($i\tends i$)
and transitive ($i\tends j,j\tends k\,\Rightarrow \, i\tends k$). Such
relations are called {\sc quasiorders}. Consider the equivalence
relation $\leftrightarrow$ on the set of events $N$:

\begin{equation}\label{e12a}
i\leftrightarrow j \:\:\, \hbox{if and only if} \:\:\,
i\tends j \: \hbox{ and } \: j\tends i
\end{equation}

\noindent and consider the quotient set

\begin{equation}\label{e13}
X \: = \: N/\leftrightarrow
\end{equation}

\noindent called {\sc finitary spacetime substitute} \cite{sorkin}
or pattern space \cite{prg}. For $x,y\in X$ introduce the relation
$x\to y$:

\begin{equation}\label{e13a}
x\to y \:\:\, \hbox{if and only if} \:\:\,
\forall i\in x, \: \forall j\in y \: \, \,i\tends j
\end{equation}

\noindent (note that the expressions like $i\in x$ make sense since
the elements of $X$ are subsets of $N$). The relation $\to$
(\ref{e13a}) on $X$ is:

\begin{itemize}
\item[(i)] reflexive: $x\to x$
\item[(ii)] transitive: $x\to y,y\to z\,\Rightarrow \, x\to z$
\item[(iii)] antisymmetric: $x\to y,y\to x\,\Rightarrow \, x=y$
\end{itemize}

The relations having these three properties are called {\sc
partial orders}. It is known (see, e.g. \cite{sorkin}) that there
is 1--1 correspondence between partial orders and topologies on
finite sets, and that the topology of the manifold can be recovered
when the number of events and observers grows to infinity
\cite{sorkin}.

\medskip

To conclude this section consider an example of finitary
substitute from \cite{g2}. Suppose there are four observers
$\obs_1,\ldots,\obs_4$ living on the circle $e^{\imath\varphi}$
whose areas of observations are:

\[
\begin{array}{ccl}
\obs_1 & \mapsto & \{-2\pi/3 < \varphi < 2\pi/3\} \cr
\obs_2 & \mapsto & \{\pi/3 < \varphi < 5\pi/3\} \cr
\obs_3 & \mapsto & \{-3\pi/4 < \varphi < 2/3\pi\}  \cr
\obs_4 & \mapsto & \{\pi/4 < \varphi < 3\pi/4\}
\end{array}
\]

Then the table of outcomes takes the form:

\begin{center}
\begin{tabular}{||c||c|c|c|c||}
\hline
Event label & $\obs_1$ & $\obs_2$ & $\obs_3$ & $\obs_4$ \cr
\hline
0 & + & -- & -- & -- \cr
$\pi/2$ & + & + & -- & + \cr
$\pi$ & -- & + & -- & -- \cr
$3\pi/2$ & + & + & + & -- \cr
\hline
\end{tabular}
\end{center}

Then the relation "$\tends$" (\ref{e12}) is already partial order:

\begin{equation}\label{e17}
\begin{array}{ccccccc}
\pi/2 & \tends & 0 &;\:& \pi/2 & \tends & \pi \cr
3\pi/2 & \tends & 0 &;\: & 3\pi/2 & \tends & \pi
\end{array}
\end{equation}

\noindent and the equivalence relation (\ref{e12a}) turns out to be
the equality. Hence the finitary substitute $X$ is the whole set of
events:

\[
x = \{\pi/2\} \; , \;
y = \{3\pi/2\} \; , \;
z = \{0\} \; , \;
w = \{\pi\}
\]

\noindent and the partial order on $X$ is:

\begin{figure}[h!t]
\unitlength2mm
\begin{center}
\begin{picture}(30,30)
\put(5,5){\circle*{1}}
\put(20,5){\circle*{1}}
\put(5,20){\circle*{1}}
\put(20,20){\circle*{1}}

\put(5,5.5){\vector(0,1){14}}
\put(20,5.5){\vector(0,1){14}}

\put(5.5,5.5){\vector(1,1){14}}
\put(19.5,5.5){\vector(-1,1){14}}

\put(2,2){\mbox{$x$}}
\put(21,2){\mbox{$y$}}
\put(2,17){\mbox{$z$}}
\put(21,17){\mbox{$w$}}

\end{picture}
\end{center}
\caption{The finitary substitute of the circle.}
\label{f17}
\end{figure}
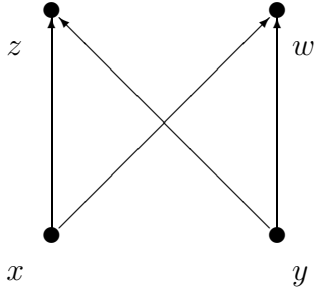

\noindent {\bf Remark.} As we have already mentioned, there is an
equivalent way to define topology in terms of converging sequences.
It worthy to mention that we use the symbol $\to$ for the
partial order (\ref{e12}) due to the following fact:

\[
x \to y \qquad \hbox{if and only if} \qquad
\lim\{x,x,\ldots,x,\ldots\} = y
\]

\section{Incidence algebras}\label{sialg}

There is no direct evidence of the compatibility of Sorkin and
histories approaches to empirical spacetime topologies. In this
section we solve this problem. We explicitly suggest the
construction which starting from the table of observations produces
an algebra whose space of primitive ideals endowed with the Rota
topology is homeomorphic to the Sorkin finitary substitute obtained
from the same table.

\medskip

As it was studied in section \ref{sentang} in order to build a
model of organized spacetime observations we need to introduce an
algebra, that is the two following objects:

\begin{itemize}
\item A linear space $H$
\item A product operation on the space $H$
\end{itemize}

\noindent somehow generated by the table of observations Tab.
\ref{tab1}, where $H$ will stand for a model of $\hh$ and the
product will capture the organization.

\medskip

As we promised, we shall deal with a linear space $H$ spanned on
the elementary propositions about topology. What could be the form
of such propositions? Each of them should involve at least two
points, since the matter of topology is just to study the mutual
positions of events. We shall choose the simplest model, namely,
that of two-point statements (a higher order situation was
considered in \cite{jmp96}). Such elementary statements were
already formulated (\ref{epropos}).

\medskip

The form of the algebra we suggest will be similar to that
introduced in \cite{dmhv}. Let $p,q$ be two events, denote by the
symbol (sic!) $\pq$ the proposition associated with this pair.
Form the linear span of all such symbols: $\algspan\{ \pq \}$ and
define the product on it:

\begin{equation}\label{epq}
\pq \cdot |r\!><\!s| \,=\, <\!q|r\!> \cdot |p\!><\!s|
\end{equation}

\noindent where \( <\!q|r\!> = \delta_{qr} \). Note the obtained
product is associative but not commutative in general.

\medskip

In order to take into account the results of the measurement (Table
\ref{tab1}) we form the linear space

\[
H \,=\, \algspan \{ \pq \hbox{ such that } p\tends q \}
\]

\noindent where $p\tends q$ (\ref{e12}) means that $p$ was
registered by a greater set of observers than $q$. To assure that
the obtained algebra can really describe an organization (in the
sense of section \ref{sentang}, we have to check that $H$ is an
algebra.

\medskip

\noindent {\bf Proposition 1.} The linear space $H$ with the product
(\ref{epq}) is associative algebra.

\medskip

\noindent {\em Proof.\/} Let $\pq$ and $|r\!><\!s|$ are in $H$,
that means $p\tends q$ and $r\tends s$. If $q\neq r$ then their
product is zero. If $q=r$ then, according to (\ref{epq}) their
product is $\pq \cdot |q\!><\!s| = |p\!><\!s|$, which is the element of $H$
since the relation $\tends$ is transitive: $p\tends q, q\tends s$
implies $p\tends s$.

\noindent {\bf Remark.} The algebras of such sort (called incidence
algebras) were introduced by Rota in \cite{rota} in a slightly
different way.

\medskip

Now let us apply the spatialization procedure described in section
\ref{sspat}. The primitive spectrum of the algebra $H$ was
calculated in \cite{drs}; it consists of all the ideals of the
form:

\begin{equation}\label{eprim}
X_s = \algspan \{ \pq :\;  \pq \neq |s\!><\!s| \}
\end{equation}

\noindent where $s$ ranges over all equivalence classes with
respect to the relation "$\leftrightarrow$" (\ref{e12a}) on $N$,
that is, events. So, at the first stage of the spatialization
procedure we already have a canonical bijection between the
elements of the primitive spectrum of the algebra $H$ and the
events in the Sorkin's discretization scheme (section \ref{sfinsub}).
In order to show the compatibility of the two schemes we have to
show that the Rota topology on the set $X_s$ is the same as that of
Sorkin.

\medskip

Let us figure out the form of the relation $\rota$ (\ref{etends})
for the suggested algebra. By the way we shall see that the
relation $\rota$ can be thought of as a sort of 'proximity'
between events. So, let $r,s$ be two events.

\medskip

\noindent {\bf Proposition 2.} Let $X_r, X_s$ be two primitive ideals.
Then $X_r\rota X_s$ if and only if $r\tends s$ and there is no
$t\neq r,s$ such that $r\tends t \tends s$.

\medskip

\noindent {\em Proof\/} will be carried out exhaustively: we shall
consider all possible cases.

\begin{itemize}
\item {\bf Case 1.} $r=s$. Consider
$X_r\cdot X_r$. To prove that this product coincides with $X_r$
recall its definition (\ref{eprim}).  Let $|a\!><\!b| \in X_r$
(that is $a\neq r$ or $b\neq r$) and prove that it can be
represented as the product of two elements from $X_r$. If $a\neq r$
then $|a\!><\!b| = |a\!><\!a|\cdot |a\!><\!b|$. If $b\neq r$ then
$|a\!><\!b| = |a\!><\!b|\cdot |b\!><\!b|$. Therefore $X_r\cdot
X_r=X_r$, and $X_r\overline{\rota} X_r$.

\item {\bf Case 2.} $r\not\tends s$. Consider an element $\pq$ from the
intersection of the ideals:

\[
X_r\cap X_s \,=\,\algspan \{ \pq :\;\pq \neq |r\!><\!r| \hbox{ and }
\pq \neq |s\!><\!s|\}
\]

\noindent and show that it belongs to the product

\[
X_r\cdot X_s   \,=\,  \algspan \{ \pq |a\!><\!b|:\; \pq \neq |r\!><\!r|
\hbox{ and } |a\!><\!b|\neq |s\!><\!s| \}
\]

\noindent If $p=q\neq r,s$ then $|p\!><\!p| = |p\!><\!p|\cdot|p\!><\!p|$. If
$p\neq q$ then $p\neq r$ or $q\neq s$ (since $r\not\tends s$). Then
$\pq = |p\!><\!p|\cdot \pq$ or $\pq = \pq\cdot |q\!><\!q|$, respectively.
Therefore $X_r\overline{\rota} X_s$.

\item {\bf Case 3.} $r\tends s$ and there is $t\neq r,s$ such that
$r\tends t \tends s$. Consider an element $\pq$ from the
intersection of the ideals:

\[
X_r\cap X_s \,=\,\algspan \{ \pq :\;\pq \neq |r\!><\!r| \hbox{ and }
\pq \neq |s\!><\!s|\}
\]

\noindent and show that it belongs to the product

\[
X_r\cdot X_s   \,=\,  \algspan \{ \pq |a\!><\!b|:\; \pq \neq |r\!><\!r|
\hbox{ and } |a\!><\!b|\neq |s\!><\!s| \}
\]

\noindent If $p=q\neq r,s$ then $|p\!><\!p| =
|p\!><\!p|\cdot|p\!><\!p|$. Let $p\neq q$ and ($p\neq r$ or $q\neq
s$), then $\pq = |p\!><\!p|\cdot \pq$ (if $p\neq r$) or $\pq = \pq
\cdot |q\!><\!q|$ (if $q\neq s$). Finally let $\pq = |r\!><\!s|$,
then $|r\!><\!s| = |r\!><\!t|\cdot |t\!><\!s|$, and we again have
$X_r\overline{\rota} X_s$.

\item {\bf Case 4.} $r\tends s$ and there is no $t\neq r,s$ such
that $r\tends t \tends s$. Let us show that the element
$|r\!><\!s|$ from the intersection of the ideals:

\[
X_r\cap X_s \,=\,\algspan \{ \pq :\;\pq \neq |r\!><\!r| \hbox{ and }
\pq \neq |s\!><\!s|\}
\]

\noindent is not an element of the product

\[
X_r\cdot X_s   \,=\,  \algspan \{ \pq |a\!><\!b|:\; \pq \neq |r\!><\!r|
\hbox{ and } |a\!><\!b|\neq |s\!><\!s| \}
\]

\noindent Suppose this is not the case, then $|r\!><\!s| = \sum
C_{atc}|a\!><\!t|\cdot|t\!><\!c|$. Multiply this equality (in $H$) by
$|r\!><\!r|$ from the left and by $|s\!><\!s|$ from the right. Then $|r\!><\!s| =
\sum C_{rts}|r\!><\!t|\cdot|t\!><\!s|$. However there is no $t$ such that
$r\tends t \tends s$, therefore this sum is zero, while $|r\!><\!s|\neq
0$.

\end{itemize}

\noindent

So we see that two primitive ideals are binded by the relation
$\rota$ if and only if they are as close as possible on the Hasse
diagram of the partial order associated with the Sorkin topology
(Case 4).

\medskip

The results of this section can be summarized in the following
theorem.

\medskip

\noindent {\bf Theorem.} {\em The Sorkin topology of a finitary
substitute coincides with the Rota topology of its incidence
algebra. }

\section*{Concluding remarks}

Initially, in the histories approach to quantum mechanics the
existence of the spacetime as a fixed manifold was presupposed
\cite{ha}. An algebraic version \cite{qlha} of this approach did
not give up this presupposition, however rendered it rudimentary.
In this paper we do the next step, and the spacetime becomes an
observable up to its combinatorial approximation.

The core of the suggested quantum scheme is the spatialization
procedure. We have realized it as close as possible to the
standard spatialization due to Gel'fand. The peculiarity
of our approach is that we impose a new topology, namely, that of
Rota (section \ref{sspat}). The reason for us to do it was that
in finite dimensional situations (which we considered as realistic
ones) the Gel'fand topology reduced to the trivial discrete one.
The suggested machinery is a bridge between the algebraic
version of the histories approach \cite{qlha} and combinatorial
models such as lattice-like discretization schemes \cite{sorkin}.

From the other side, beyond the histories approach an algebraic
construction merging finitary substitutes into the quantum-like
environment was already carried out by the 'poseteers group' (the
term introduced by F.Lizzi) in \cite{g2,g1}. The comparison of the
two constructions is in Table \ref{tabc}.

\begin{table}[h!t]
\begin{tabular}{p{0.35\textwidth}%
@{\hspace{0.1\textwidth}}p{0.35\textwidth}}
\hline \\
{\sc Poseteers' approach} & {\sc Incidence algebras}\\
&\\
\multicolumn{2}{c}{\bf The algebras}\\
&\\
$C^*$-algebras of infinite dimensions &
finite dimensional algebras with no involution \\
&\\
\multicolumn{2}{c}{\bf The points}\\
&\\
kernels of irreducible \mbox{$*$-re}presentations &
kernels of irreducible representations \\
&\\
\multicolumn{2}{c}{\bf The topology}\\
&\\
Jacobson topology & Rota topology \\
&\\
\hline \\
\end{tabular}
\caption{The comparison of the two algebraic schemes.}
\label{tabc}
\end{table}

Note that the incidence algebras are not semisimple. At first sight
this seems to be an essential drawback of the theory bringing it
far from the existing quantum constructions. However, in the case
of finite dimension the Jacobson topology will always be
discrete.  From the other side, the lack of semisimplicity makes it
possible to develop differential calculi on finitary models, which
may be considered as a link between the poseteers' approach and the
formalism of discrete differential manifolds \cite{dmhv}.

In \cite{dmhv} finite dimensional semisimple commutative algebras
are studied and a differential structure is built in terms of
moduli of differential forms (being the conjugate to the module of
derivations in the classical case), while there is no nonzero
derivations in the algebras themselves. Contrarily, the incidence
algebras possess derivations which makes it possible to introduce
tensor calculi based on the notion of duality \cite{dv}. Moreover, since all their derivatives are inner, they already contain vectors (for details the reader is referred to \cite{ricomm}).

Finally, we dwell on the algebra of symbols $\pq$ we introduced in
section \ref{sialg}. Irrespective to the particular form of the
organizing algebra (like the incidence algebra in our case) we can
always write down expressions like

\[ \large
\sum_{q_1,\ldots,q_n} |p\!><\!q_1|\circ |q_1\!><\!q_2|\circ
\ldots \circ |q_n\!><\!r|
\]

\noindent where the operation "$\circ$" is a multiplication
generating a particular organization of the team (section
\ref{sentang}), and $q_i,q_{i+1}$ are neighbor (in the sense of
Rota topology) events. So, this expression can be thought of as the
sum over trajectories.

\medskip

One of the authors (RRZ) expresses his gratitude to prof. G.Landi
who organized his visit to the University of Trieste (supported by
the Italian National Research Council) and conveyed many important
ideas. Stimulating discussions with P.M.Ha\-jac,
S.V.Kra\-s\-ni\-kov, and Te\-t\-sue Ma\-s\-u\-da are appreciated.

The work was supported by the RFFI research grant (96--02--19528).
One of us (R.R.Z.) acknowledges the financial support from the
Soros foundation (grant A98-42) and the research grant 97--14.3--62
"Universities of Russia".

\end{document}